\documentclass[a4paper]{article}
\usepackage[german, english]{babel}
\usepackage{a4wide}
\usepackage{amsmath}
\usepackage{amssymb}
\usepackage{ifthen}
\usepackage{epsfig}
\usepackage[hang,nooneline]{subfigure}
\newcounter{fig}

\begin{document}

\title{\bf Charged Boson Stars and Black Holes}
\vspace{1.5truecm}
\author{
{\bf Burkhard Kleihaus, Jutta Kunz}\\
Institut f\"ur  Physik, Universit\"at Oldenburg, Postfach 2503\\
D-26111 Oldenburg, Germany\\
\hspace{0.1cm}\\
{\bf Claus L\"ammerzahl, Meike List}\\
ZARM, Universit\"at Bremen, Am Fallturm\\
D-28359 Bremen, Germany
}

\vspace{1.5truecm}

\date{\today}

\maketitle
\vspace{1.0truecm}

\begin{abstract}
We consider boson stars and black holes in scalar electrodynamics
with a V-shaped scalar potential.
The boson stars come in two types, having either
ball-like or shell-like charge density.
We analyze the properties of these solutions
and determine their domains of existence.
When mass and charge become equal, the space-times
develop a throat.
The shell-like solutions need not be globally regular,
but may possess a horizon.
The space-times then consist of
a Schwarzschild-type black hole in the interior,
surrounded by a shell of charged matter,
and thus a Reissner-Nordstr\"om-type space-time in the exterior.
These solutions violate black hole uniqueness.
The mass of the black hole solutions is related 
to the mass of the regular shell-like solutions 
by a mass formula of the type
first obtained within the isolated horizon framework.
\end{abstract}

\section{Introduction}

$Q$-balls represent stationary localized solutions 
of a complex scalar field theory with
a suitable self-interaction in flat space
\cite{Friedberg:1976me,Coleman:1985ki}.
The global phase invariance of the scalar field theory
is associated with a conserved charge $Q$ \cite{Friedberg:1976me},
which represents the electromagnetic charge,
once the theory is promoted to a gauge theory.

The simplest type of $Q$-balls is spherically symmetric.
These possess a finite mass and charge, but carry no angular momentum.
Considering their mass as a function of their charge,
there are two branches of $Q$-balls, merging and ending at a cusp,
where mass and charge assume their minimal values \cite{Friedberg:1976me}.

Recently, a new type of scalar potential for $Q$-balls was considered,
leading to the signum-Gordan equation for the scalar field
\cite{Arodz:2008jk,Arodz:2008nm}.
This potential gives rise to spatially compact $Q$-balls, 
where the scalar field vanishes identically outside a critical
radius $r_o$ \cite{Arodz:2008jk}.
When coupled to electromagnetism, a new type of solution appears,
$Q$-shells \cite{Arodz:2008nm}.
In $Q$-shells the scalar field vanishes identically both inside
a critical radius $r_i$ and outside a critical radius $r_o$,
thus forming a finite shell $r_i < r < r_o$ of charged matter.

When gravity is coupled to $Q$-balls, boson stars arise,
representing globally regular self-gravitating solutions
\cite{Lee:1991ax,Jetzer:1991jr,Mielke:2000mh,Schunck:2003kk}.
The presence of gravity has a crucial influence
on the domain of existence of the classical solutions.
Instead of only two branches of solutions joint at a single cusp, 
the boson stars exhibit an intricate cusp structure,
where mass and charge oscillate endlessly.
For black holes with scalar fields, on the other hand, 
a number of theorems exist, which exclude their existence
under a large variety of conditions 
\cite{Bekenstein:1971hc,Bekenstein:1995un,Mayo:1996mv}.

Here we consider the effect of gravity on the $Q$-balls and
$Q$-shells of the signum-Gordon model coupled to a Maxwell field.
We construct these charged boson stars and gravitating $Q$-shells
and analyze their properties and their domains of existence.
We observe that at certain critical values of the mass and charge,
the space-times form a throat at the (outer) radius $r_o$, 
rendering the respective exterior space-time
an exterior extremal Reissner-Nordstr\"om space-time.

Moreover, we show that in this model
the black hole theorems can be elluded,
that forbid black holes with scalar hair.
Indeed, the gravitating $Q$-shells can be endowed with a horizon $r_H$
in the interior region $0< r_H < r_i$,
where the scalar field vanishes and the gauge potential is constant.
This Schwarzschild-type black hole in the interior
is surrounded by a shell of charged matter, $r_i<r<r_o$,
leading to an exterior space-time $r_o<r<\infty$ of
Reissner-Nordstr\"om type. 
We analyze the global and horizon properties of these black holes 
within $Q$-shells, and present a mass relation.
When a throat develops at the outer radius $r_o$,
which renders the respective exterior space-time
an exterior extremal Reissner-Nordstr\"om space-time,
the temperature of the horizon $r_H$ tends to zero.

In section 2 we recall the action and ans\"atze for the fields.
We present the gravitating $Q$-balls and $Q$-shells in section 3 and 4, 
respectively, and the black holes within $Q$-shells in section 5.
We end with a conclusion and an outlook.

\section{Action}

We consider the action of a self-interacting complex scalar field
$\Phi$ coupled to a U(1) gauge field and to Einstein gravity
\begin{equation}
S=\int \left[ \frac{R}{16\pi G}
   - \frac{1}{4} F^{\mu\nu} F_{\mu\nu}
   -  \left( D_\mu \Phi \right)^* \left( D^\mu \Phi \right)
 - U( \left| \Phi \right|) 
 \right] \sqrt{-g} d^4x
 , \label{action}
\end{equation}
with field strength tensor
\begin{equation}
F_{\mu\nu} = \partial_\mu A_\nu - \partial_\nu A_\mu 
 , \label{action2}
\end{equation}
covariant derivative
\begin{equation}
D_\mu \Phi = \partial_\mu \Phi + i e A_\mu \Phi
 , \label{action3}
\end{equation}
curvature scalar $R$, Newton's constant $G$, 
gauge coupling constant $e$,
and the asterisk denotes complex conjugation.
The scalar potential $U$ is chosen as
\begin{equation}
U(|\Phi|) =  \lambda  |\Phi| 
 . \label{U} \end{equation} 

Variation of the action with respect to the metric and the matter fields
leads, respectively, to the Einstein equations
\begin{equation}
G_{\mu\nu}= R_{\mu\nu}-\frac{1}{2}g_{\mu\nu}R = 8\pi G T_{\mu\nu}
\  \label{ee} \end{equation}
with stress-energy tensor
\begin{eqnarray}
T_{\mu\nu} &=& g_{\mu\nu}{L}_M
-2 \frac{\partial {L}_M}{\partial g^{\mu\nu}}
=
    ( F_{\mu\alpha} F_{\nu\beta} g^{\alpha\beta}
   -\frac{1}{4} g_{\mu\nu} F_{\alpha\beta} F^{\alpha\beta})
\nonumber\\
&-& 
   \frac{1}{2} g_{\mu\nu} \left(     (D_\alpha \Phi)^* (D_\beta \Phi)
  + (D_\beta \Phi)^* (D_\alpha \Phi)    \right) g^{\alpha\beta}
  + (D_\mu \Phi)^* (D_\nu \Phi) + (D_\nu \Phi)^* (D_\mu \Phi)
\nonumber\\
&-& 
    {\lambda} g_{\mu\nu}  |\Phi| 
 , \label{tmunu}
\end{eqnarray}
and the matter field equations,
\begin{eqnarray}
& & \partial_\mu \left ( \sqrt{-g} F^{\mu\nu} \right) =
   \sqrt{-g} e \Phi^* D^\nu \Phi 
\label{feqA} \end{eqnarray}
\begin{eqnarray}
& &D_\mu D^\mu \Phi = - \frac{\lambda}{2} \frac{\Phi}{|\Phi|}
 . \label{feqH} \end{eqnarray}

To construct static spherically symmetric solutions
we employ Schwarz\-schild-like coordinates and adopt
the spherically symmetric metric
\begin{equation}
ds^2=g_{\mu\nu}dx^\mu dx^\nu=
  -A^2N dt^2 + N^{-1} dr^2 + r^2 (d\theta^2 + \sin^2\theta d\phi^2)
 , \end{equation}
with
\begin{equation}
N=1-\frac{2m(r)}{r}
 . \end{equation}

For solutions with vanishing magnetic field
the Ansatz for the matter fields has the form
\begin{equation}
 \Phi = \phi(r) e^{i \omega t}
 , \label{phi} \end{equation}
\begin{equation}
 A_\mu d x^\mu = A_0(r) dt
 . \label{A_0} \end{equation}
 
For notational simplicity, we 
introduce new coupling constants \cite{Arodz:2008nm}
\begin{equation}
\alpha^2 = a = 4\pi G \frac{\beta^{1/3}}{e^2},
\ \ \
\beta = \frac{\lambda e}{\sqrt{2}} 
 , \label{constants} \end{equation}
and redefine the matter field functions,
\begin{equation}
h(r) = \sqrt{2}e \phi(r) ,
\ \ \
b(r) = \omega + e A_0(r)
 . \label{functions} \end{equation}
The latter corresponds to
performing a gauge transformation to make the scalar field real
and absorbing the frequency $\omega$ of the scalar field into the gauge
transformed vector potential.
Note, that the parameter $\beta$ can be removed by rescaling 
and will therefore be set to one
\cite{Arodz:2008nm}.
Thus the only parameter left is the gravitational coupling $\alpha$.

Let us now specify the boundary conditions for the 
metric and matter functions.
For the metric function $A$ we adopt
\begin{equation}
A(r_o)=1
\ , \end{equation}
where $r_o$ is the outer radius,
thus fixing the time coordinate. 
For the mass function $m(r)$ we require for globally regular ball-like
boson star solutions
\begin{equation}
m(0)=0 
, \end{equation}
for globally regular shell-like solutions 
\begin{equation}
m(r_i)=0 
, \end{equation}
where $r_i$ is the inner radius of the shell,
and for black hole solutions
\begin{equation}
m(r_H)= \frac{r_H}{2}
  \end{equation}
where $r_H < r_i$ denotes the horizon.

To be able to specify more than four boundary conditions 
for the matter functions we introduce one or more
auxiliary variables.
For globally regular boson star solutions we require
at the origin and at the outer radius $r_o$
the conditions
\begin{equation}
b'(0)=0 , \ \ \ h'(0)=0 , \ \ \ h(r_o)=0 , \ \ \ h'(r_o)=0
, \end{equation}
where the prime denotes differentiation with respect to $r$.
In order to choose also the value of $b(0)$ as a boundary condition,
we make the outer radius $r_o$ an auxiliary (constant) variable,
and thus add the differential equation $r_o\,'=0$,
without imposing a boundary condition.

For globally regular shell solutions as well as for black holes
we require at the inner radius $r_i$ and at the outer radius $r_o$
the conditions
\begin{equation}
b'(r_i)=0 , \ \ \ h(r_i)=0 , \ \ \ h'(r_i)=0 , \ \ \ h(r_o)=0 , \ \ \ h'(r_o)=0
 . \end{equation}
In order to choose also the value of $b(r_i)$ as a boundary condition,
we now also make the ratio of inner and outer radius $r_i/r_o$
an auxiliary (constant) variable.
Alternatively to demanding a certain value for $b(r_i)$,
we may also specify the value of the electric charge $Q$.

The charge $Q$ of the solutions is associated with the conserved current
\begin{eqnarray}
j^{\mu} & = &  - i \left( \Phi^* D^{\mu} \Phi
 - \Phi D^{\mu}\Phi ^* \right) \ , \ \ \
j^{\mu} _{\ ; \, \mu}  =  0 \ ,
\end{eqnarray}
i.e., the charge is obtained as the spatial integral
\begin{equation}
Q = -i \int j^0 \sqrt{-g} dr d\theta d\varphi
. \end{equation}

The mass $M$ of the stationary asymptotically flat space-times
is obtained from the corresponding Komar expression.
For globally regular space-times like boson stars
and shells of boson matter the mass is given by
\begin{equation}
M = 
 \frac{1}{{4\pi G}} \int_{\Sigma}
 R_{\mu\nu}n^\mu\xi^\nu dV
 , \label{komarM1}
\end{equation}
where $\Sigma$ denotes an asymptotically flat spacelike hypersurface,
$n^\mu$ is normal to $\Sigma$ with $n_\mu n^\mu = -1$,
$dV$ is the natural volume element on $\Sigma$,
and $\xi$ denotes an asymptotically timelike Killing vector field
\cite{wald}.
Replacing the Ricci tensor via the Einstein equations by the
stress-energy tensor yields
\begin{equation}
M = 
  \, 2 \int_{\Sigma} \left(  T_{\mu \nu}
-\frac{1}{2} \, g_{\mu\nu} \, T_{\gamma}^{\ \gamma}
 \right) n^{\mu }\xi^{\nu} dV
 . \label{komarM2}
\end{equation}

For black hole space-times the corresponding Komar expression is given by
\begin{equation}
M = M_H +
  \, 2 \int_{\Sigma} \left(  T_{\mu \nu}
-\frac{1}{2} \, g_{\mu\nu} \, T_{\gamma}^{\ \gamma}
 \right) n^{\mu }\xi^{\nu} dV 
 , \label{komarM3}
\end{equation}
where $M_H$ is the horizon mass of the black hole.
The mass of all (gravitating) solutions
can be directly obtained from the asymptotic form of their metric.
In the units employed, we find
\begin{equation}
M= \frac{1}{\alpha^2}\, \lim_{r \rightarrow \infty} m(r)
 . \label{mass} \end{equation}

\section{Boson Stars}

\begin{figure}[p]
\begin{center}
\vspace{-1.5cm}
\mbox{\hspace{-1.5cm}
\subfigure[][]{
\includegraphics[height=.27\textheight, angle =0]{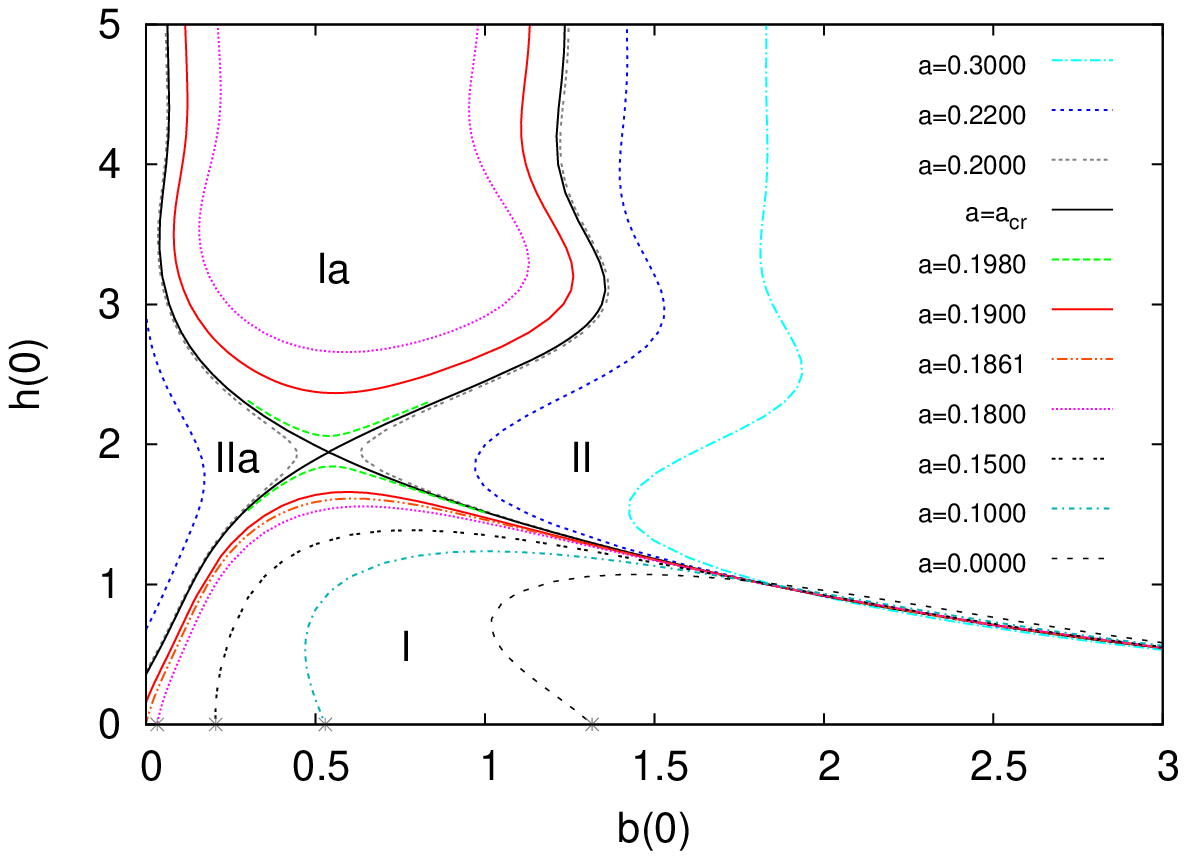}
\label{phasediag}
}
\subfigure[][]{\hspace{-0.5cm}
\includegraphics[height=.27\textheight, angle =0]{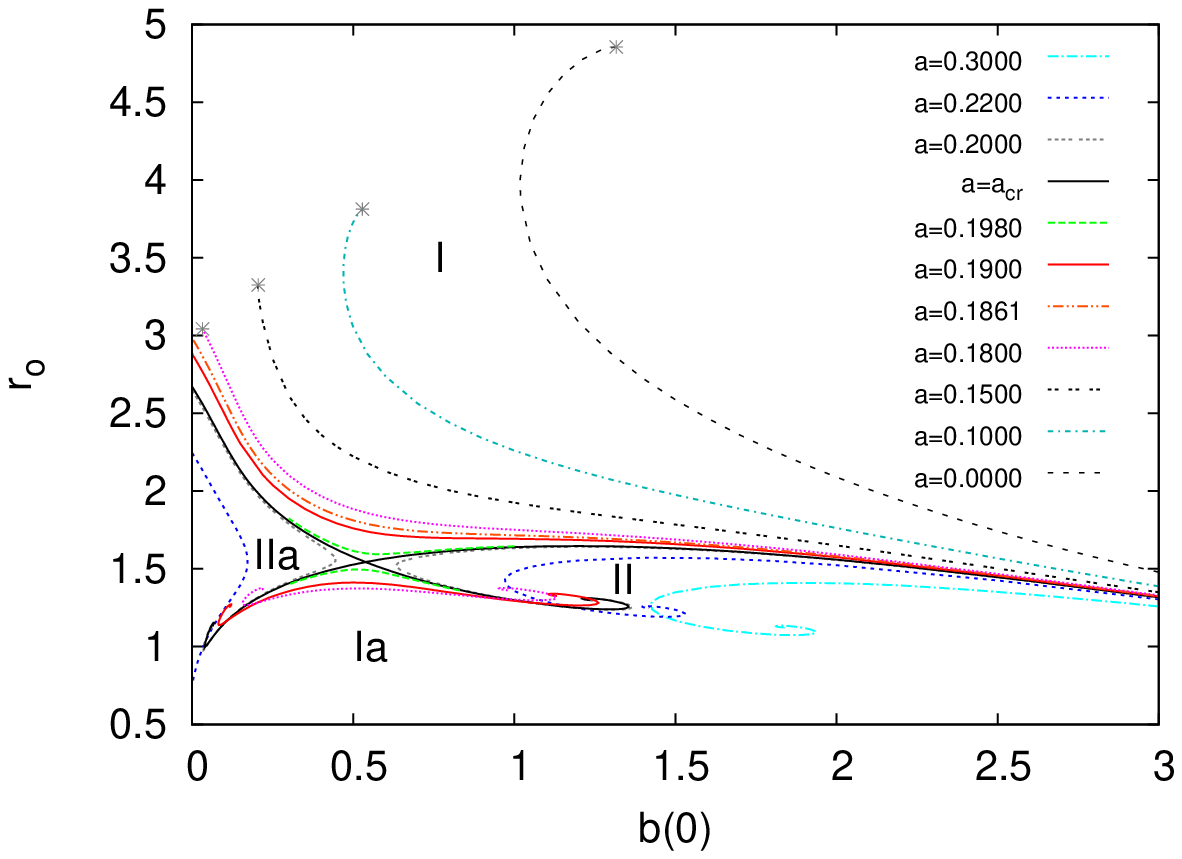}
\label{QB_r0_vs_b0}
}
}
\vspace{-0.5cm}
\mbox{\hspace{-1.5cm}
\subfigure[][]{
\includegraphics[height=.27\textheight, angle =0]{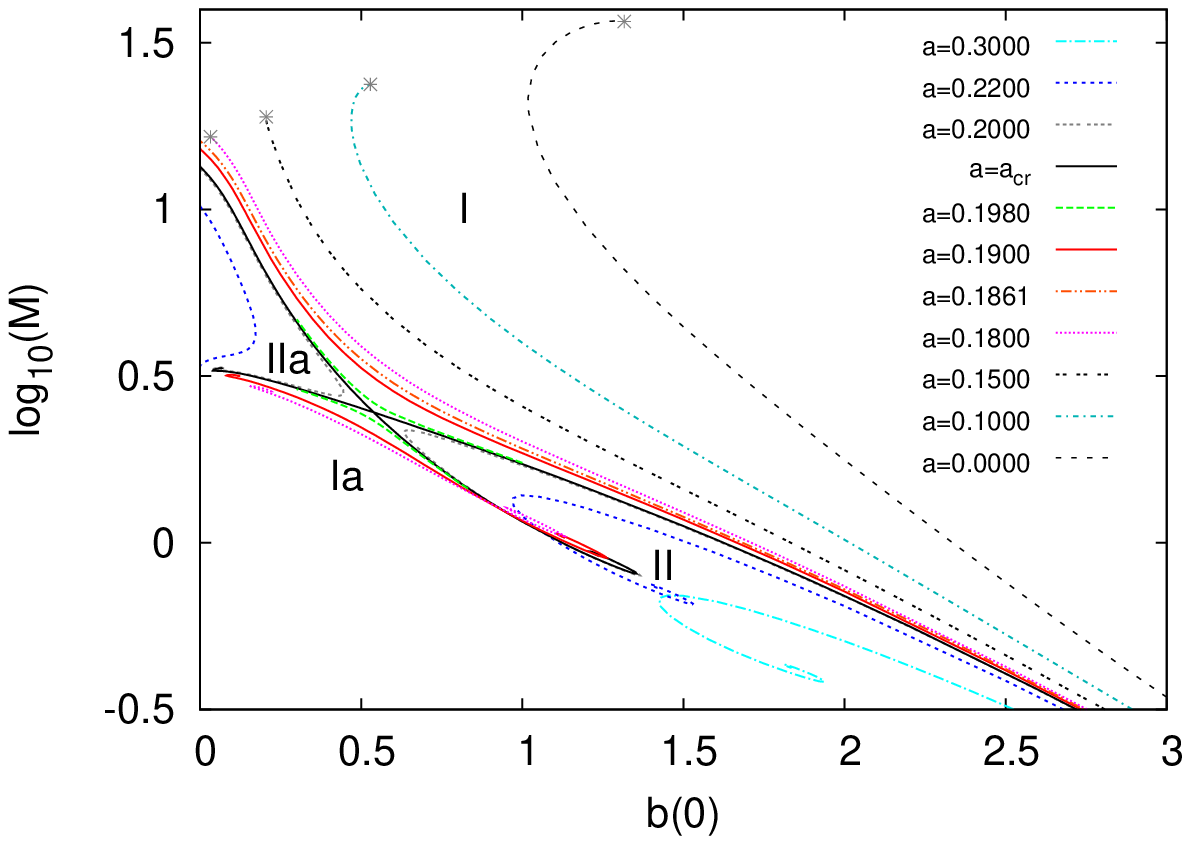}
\label{QB_M_vs_b0}
}
\subfigure[][]{\hspace{-0.5cm}
\includegraphics[height=.27\textheight, angle =0]{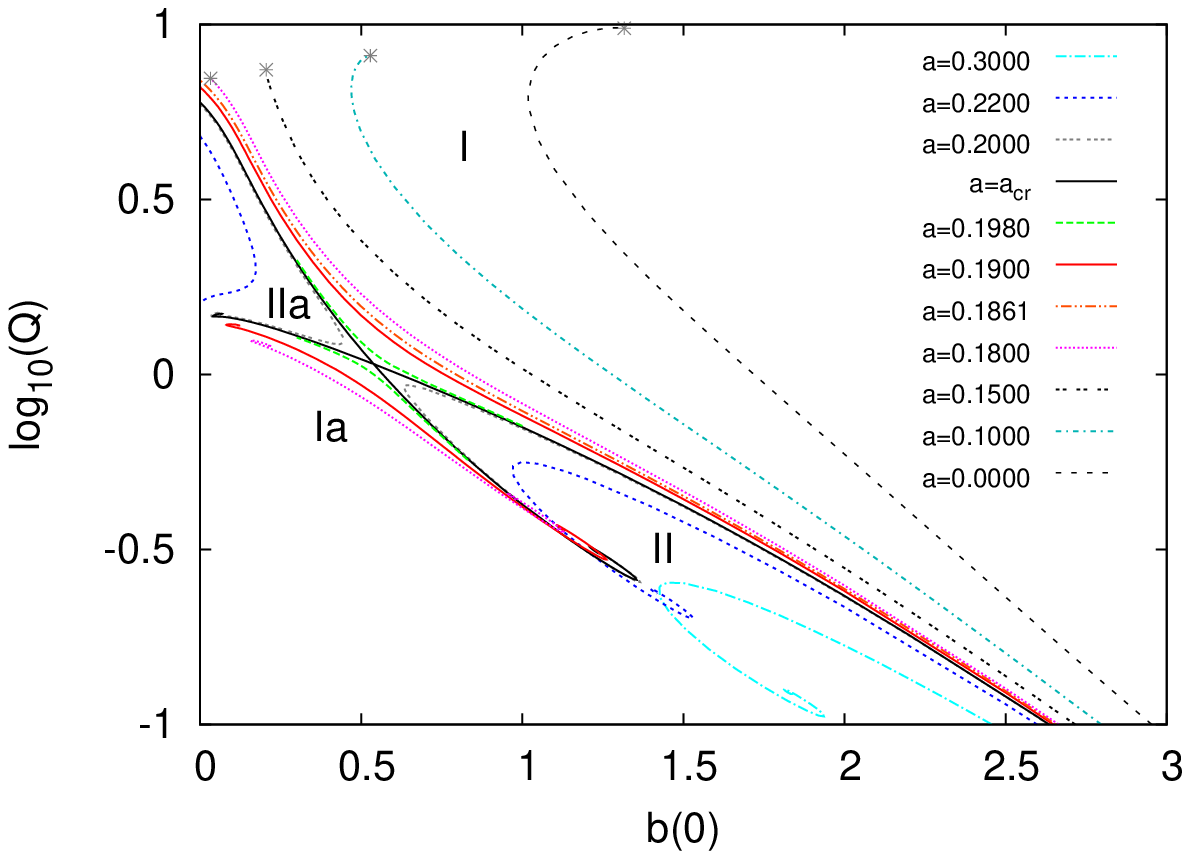}
\label{QB_Q_vs_b0}
}
}
\vspace{-0.5cm}
\mbox{\hspace{-1.5cm}
\subfigure[][]{
\includegraphics[height=.27\textheight, angle =0]{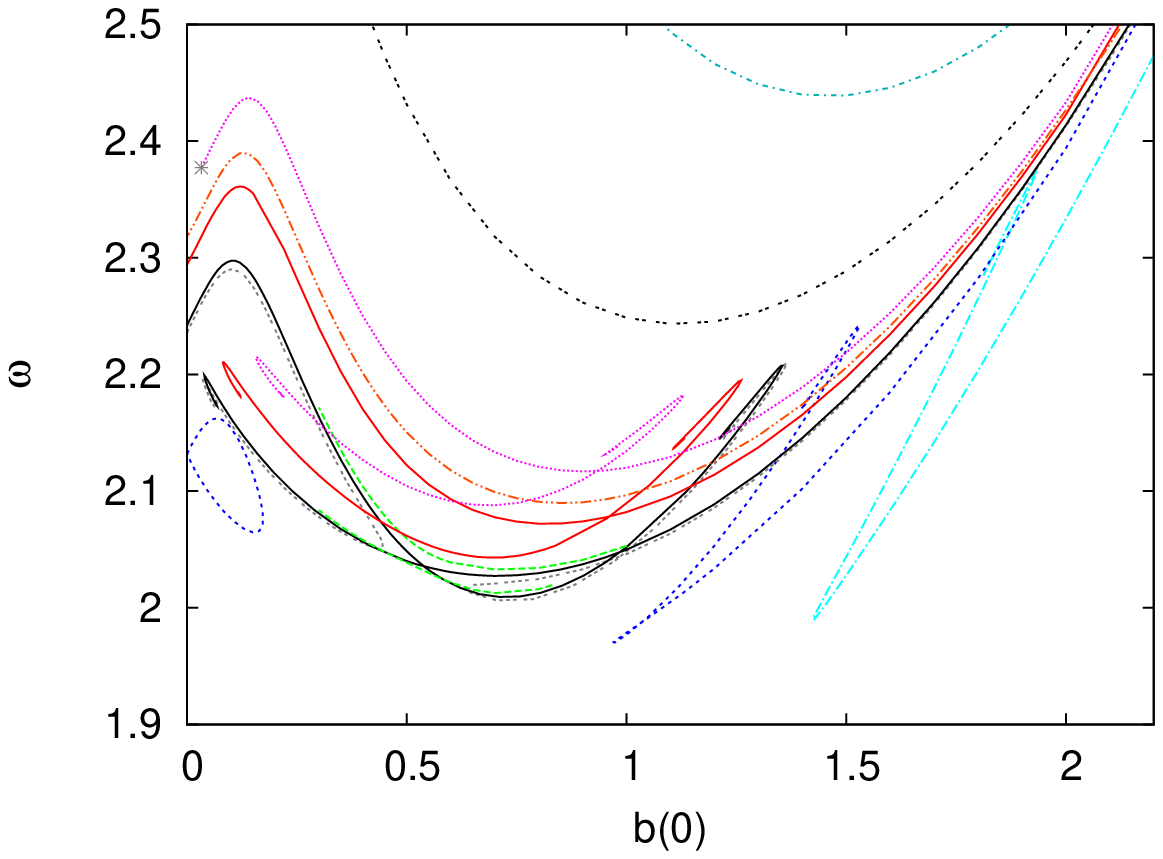}
\label{QB_om_vs_b0}
}
\subfigure[][]{\hspace{-0.5cm}
\includegraphics[height=.27\textheight, angle =0]{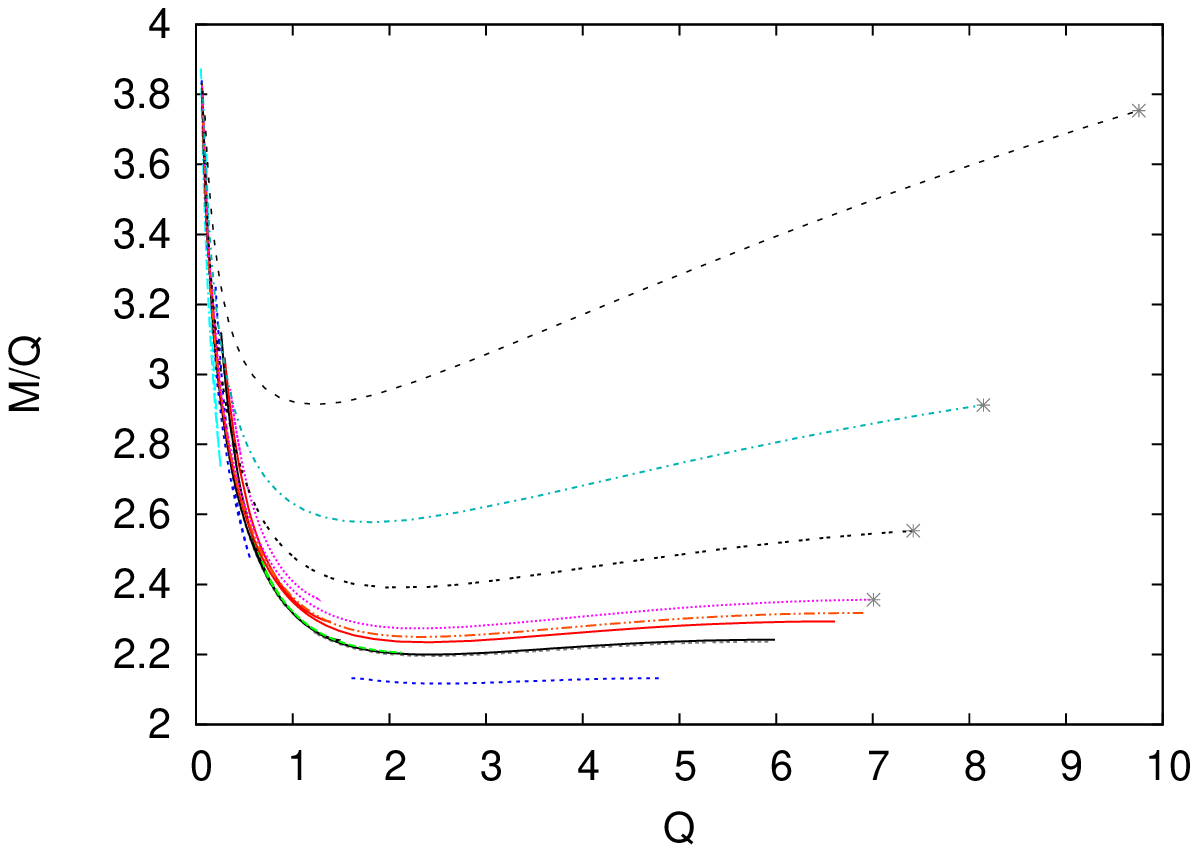}
\label{QB_Mvs_Q}
}
}
\end{center}
\caption{Properties of the boson star solutions shown versus $b(0)$,
the value of the gauge field function $b(r)$ at the origin:
(a) $h(0)$, the value of the scalar field function $h(r)$ at the origin,
forms with $b(0)$ the phase diagram of the solutions;
(b) $r_o$ indicates the size of the solutions;
(c) and (d) exhibit logarithmically
the mass $M$ and the charge $Q$, respectively;
and (e) shows the value $b(\infty)$ of the gauge field function
$b(r)$ at infinity (which corresponds to the value
of the scalar field frequency $\omega$).
In (f) the ratio of mass and charge $M/Q$ is exhibited versus 
the charge $Q$.
Note that $a=\alpha^2$, and the asterisks mark the transition
points from boson stars ($Q$-balls) to $Q$-shells.
\label{bosonstar}
}
\end{figure}

We first address the set of globally regular 
ball-like boson star solutions,
obtained by coupling the $Q$-ball solutions \cite{Arodz:2008nm} to gravity.
We illustrate the physical properties of these solutions
in Fig.~\ref{bosonstar}.

Fig.~\ref{phasediag} represents the phase diagram for the boson star
solutions. Here the sets of solutions for a sequence of values of the
gravitational coupling $\alpha$ are exhibited in terms of
the values of the matter field functions at the center of the solutions,
i.e., the value of the scalar field function $h(0)$ 
and the value of the gauge field function $b(0)$.
The figure consists of four distinct regions,
which we have labelled I, Ia, II and IIa.
For $\alpha=0$, the $Q$-ball solutions form a continuous set,
represented by a single curve in the lower part of the figure,
denoted as region I.
These non-gravitating solutions are
bounded by some maximal value of $h(0)$, by some minimal value of $b(0)$,
and by the bifurcation point with the shell-like solutions,
where $h(0)$ reaches zero.

As the gravitational coupling constant $\alpha$ is increased from zero, 
the resulting sets of solutions smoothly deform and fill region I.
The maximal value of $h(0)$ 
of these sets of solutions (for each value of $\alpha$ given by 
a single continuous curve) then increases,
the corresponding minimal value of $b(0)$ decreases, 
and the bifurcation point with the shell-like solutions decreases as well.
This decrease of the bifurcation point continues until 
zero is reached at the critical value of 
the gravitational coupling constant, $\alpha_{sh} \approx 0.1861$.
Beyond $\alpha_{sh}$ no shell-like solutions exist.
However,
the set of solutions continues to deform smoothly with increasing $\alpha$.

This smooth evolution continues, until a second 
critical value of the gravitational coupling constant $\alpha$ is reached,
$\alpha_{cr} \approx 0.1989$.
Here a bifurcation with a second set of solutions is encountered,
which constitute the boundary of region Ia.
This second type of solutions is present for each finite value of
$\alpha \le \alpha_{cr}$. 
For this type of solutions $h(0)$ has a minimal value for fixed $\alpha$,
which decreases with increasing $\alpha$,
until at $\alpha_{cr}$ the bifurcation is reached.

At $\alpha_{cr}$ the set I and set Ia
solutions touch and bifurcate. 
For $\alpha > \alpha_{cr}$ 
they then split into a right and left set of solutions,
forming regions II and IIa, respectively.
The solutions in region II correspond to the larger values of $b(0)$,
while the solutions in region IIa are restricted to the
smaller values of $b(0)$.
With increasing $\alpha$, the sets of solutions in region IIa
move towards smaller values of $b(0)$,
possibly disappearing at some critical value of the gravitational coupling,
whereas the sets of solutions in region II
move towards larger values of $b(0)$.

Fig.~\ref{QB_r0_vs_b0} shows the outer radius $r_o$ for these sets of solutions,
and thus the size of the corresponding boson stars.
Clearly, the biggest size for a given $\alpha \le \alpha_{cr}$
is always reached in region I at the bifurcation point
with the shell-like solutions.
The oscillations of the gauge field value $b(0)$ 
with increasing scalar field value $h(0)$ seen in region II in
Fig.~\ref{phasediag}
are reflected in the spirals
formed by the outer radius $r_o$ in region II in Fig.~\ref{QB_r0_vs_b0}.
They are also present in regions IIa and Ia, whenever
the gauge field value $b(0)$ exhibits oscillations.

The mass $M$ and the charge $Q$ of these sets of boson star solutions
are exhibited in Figs.~\ref{QB_M_vs_b0} and \ref{QB_Q_vs_b0}.
Both show a very similar pattern.
Again, the biggest mass and charge for a given $\alpha \le \alpha_{cr}$
are reached in region I at the respective bifurcation point
with the shell-like solutions,
while the oscillations of $b(0)$ seen in regions Ia, II and IIa
lead to spiral patterns for the mass and charge.

The corresponding family of curves for the asymptotic
value $b(\infty)$ of the gauge field function $b(r)$ at infinity
(which can be indentified with the value of the
scalar field frequency $\omega$ in the gauge,
where the gauge field vanishes at infinity)
is exhibited in Fig.~\ref{QB_om_vs_b0}.
Here the overall pattern is different, but spirals occur as well.
Finally, in Fig.~\ref{QB_Mvs_Q} we exhibit
the ratio of mass and charge $M/Q$ versus $Q$.
We observe a linear increase of $M/Q$ with $Q$ for the larger values
of $Q$ in regions I and IIa, where
the slope decreases with increasing $\alpha$,
making $M/Q$ almost constant for larger values of 
$\alpha$ (e.g.,~$\alpha=0.22$).

While the occurrence of spirals is a typical feature of 
boson star solutions \cite{Friedberg:1976me}, 
the present sets of solutions exhibit a for boson stars new phenomenon,
namely the formation of throats.
As a throat is formed,
the minimum of the metric function $N(r)$ tends to zero,
and the zero is reached precisely at the outer radius $r_o$.
At the same time the metric function $A(r)$ tends to a step
function, that vanishes inside $r_o$, and assumes the asymptotic value
$A(r)=1$ outside $r_o$.
(In Fig.~\ref{fun} the functions close to throat formation
are exhibited in the case of black holes.)

The space-time for $r \ge r_o$ then corresponds to the exterior
space-time of an extremal Reissner-Nordstr\"om (RN) black hole.
Indeed, there the metric function $N(r)$ can be expressed as
\begin{equation}
N(r) = 1 - \frac{2 \alpha^2 M}{r} + \frac{\alpha^2 Q^2}{r^2} 
 = \left( 1 - \frac{\alpha Q}{r} \right)^2 
 , \label{RN} \end{equation}
i.e., $r_H = r_o= \alpha^2 M = \alpha Q$ for the extremal RN solution
(in the units employed).
As seen in Fig.~\ref{QB_Mvs_Q},
this relation is precisely satisfied, when $b(0) \rightarrow 0$.
Thus a throat is formed, when in a set of solutions the value $b(0)$ 
of the gauge field function tends to zero.
In fact, the function $b(r)$ then tends to zero in the whole region
$r< r_o$, and its derivative $b'(r)$ does so as well.
However, at $r_o$ the derivative $b'(r)$ jumps to a finite
value, necessary for the 
Coulomb fall-off of a solution with charge $Q$.

Finally we note, that the sets of boson star solutions
with fixed gravitational coupling constant $\alpha$ 
satisfy a mass relation.
This relation is based on the observation, that
\begin{equation}
d M = b(\infty) d Q 
 , \label{Mreg2} \end{equation}
shown to hold for the regular solutions in flat space \cite{Arodz:2008nm}.
Since (\ref{Mreg2}) continues to hold for the gravitating solutions,
integration 
yields the mass relation
\begin{equation}
M_2 =  M_1 + M_Q =
 M_1 + \int_{Q_1}^{Q_2} b(\infty) d Q
 , \label{Mreg} \end{equation}
where the mass $M_2$ of a regular solution with charge $Q_2$
is obtained 
by integrating from any regular solution $M_1$ with charge $Q_1$
along the curve of intermediate solutions of the set.

\section{Gravitating $Q$-Shells}
 
\begin{figure}[t!]
\begin{center}
\mbox{\hspace{-0.5cm}
\subfigure[][]{\hspace{-1.0cm}
\includegraphics[height=.27\textheight, angle =0]{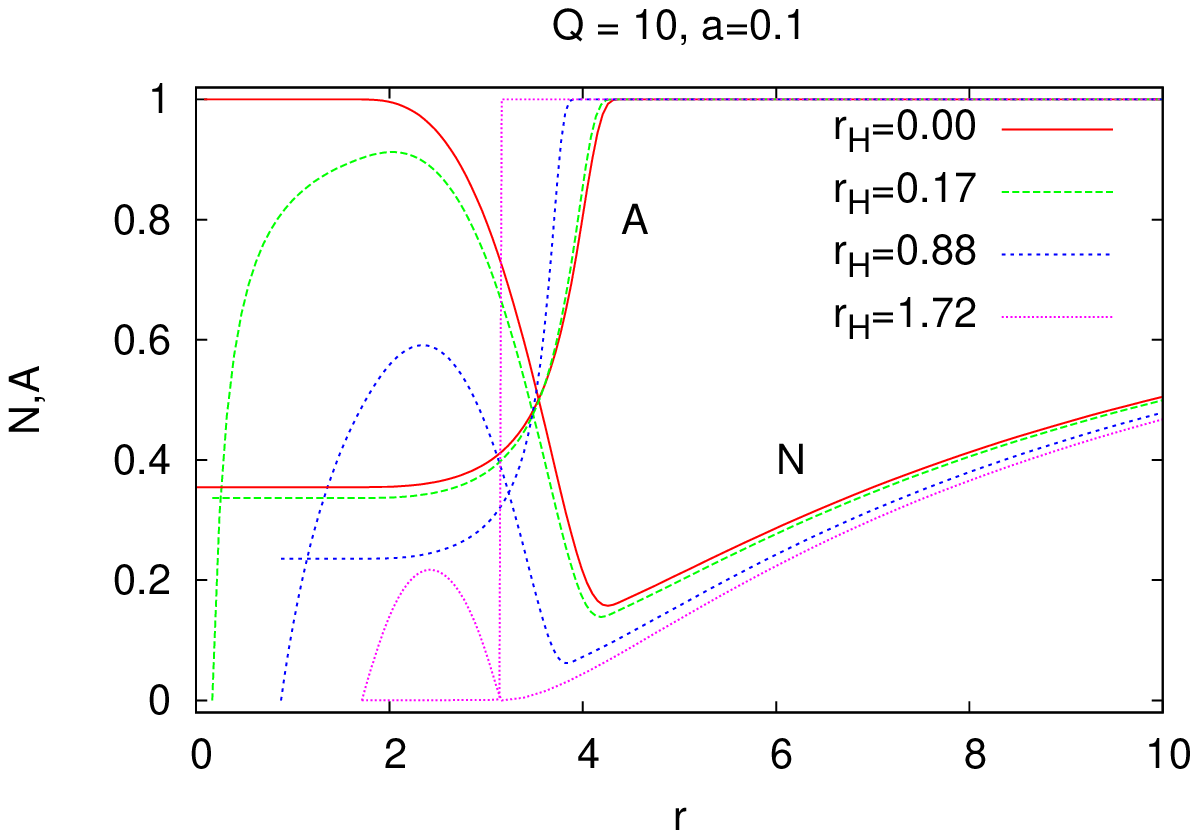}
\label{fun1}
}
\subfigure[][]{\hspace{-0.5cm}
\includegraphics[height=.27\textheight, angle =0]{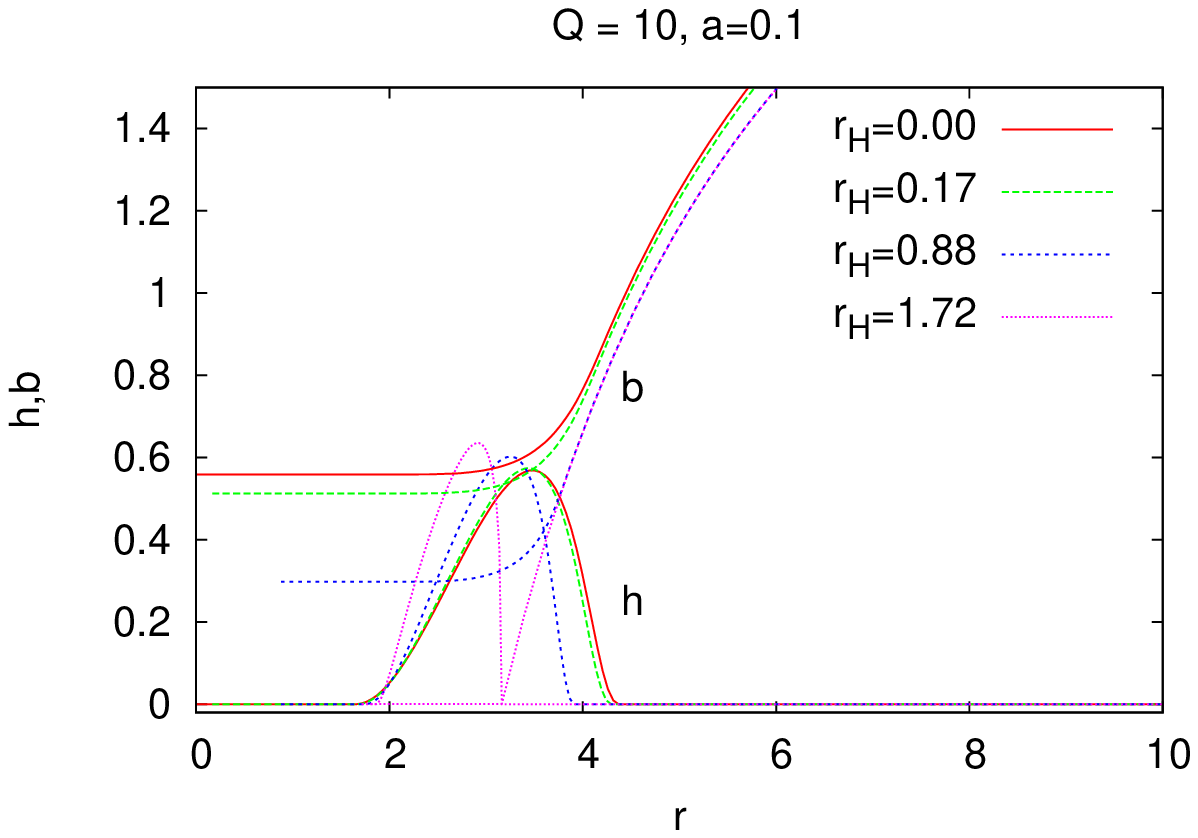}
\label{fun2}
}
}
\end{center}
\caption{Functions of the gravitating $Q$-shell solutions 
shown versus the radial coordinate $r$ for $Q=10$ and $\alpha^2=0.1$:
(a) metric functions $A(r)$ and $N(r)$;
(b) matter functions $h(r)$ and $b(r)$.
Also shown are the corresponding functions for
black hole solutions with several horizon radii $r_H$.
The largest $r_H$ is close to the critical
value, where the throat is formed.
\label{fun}
}
\end{figure}

\begin{figure}[t!]
\begin{center}
\mbox{\hspace{-0.5cm}
\subfigure[][]{\hspace{-1.0cm}
\includegraphics[height=.27\textheight, angle =0]{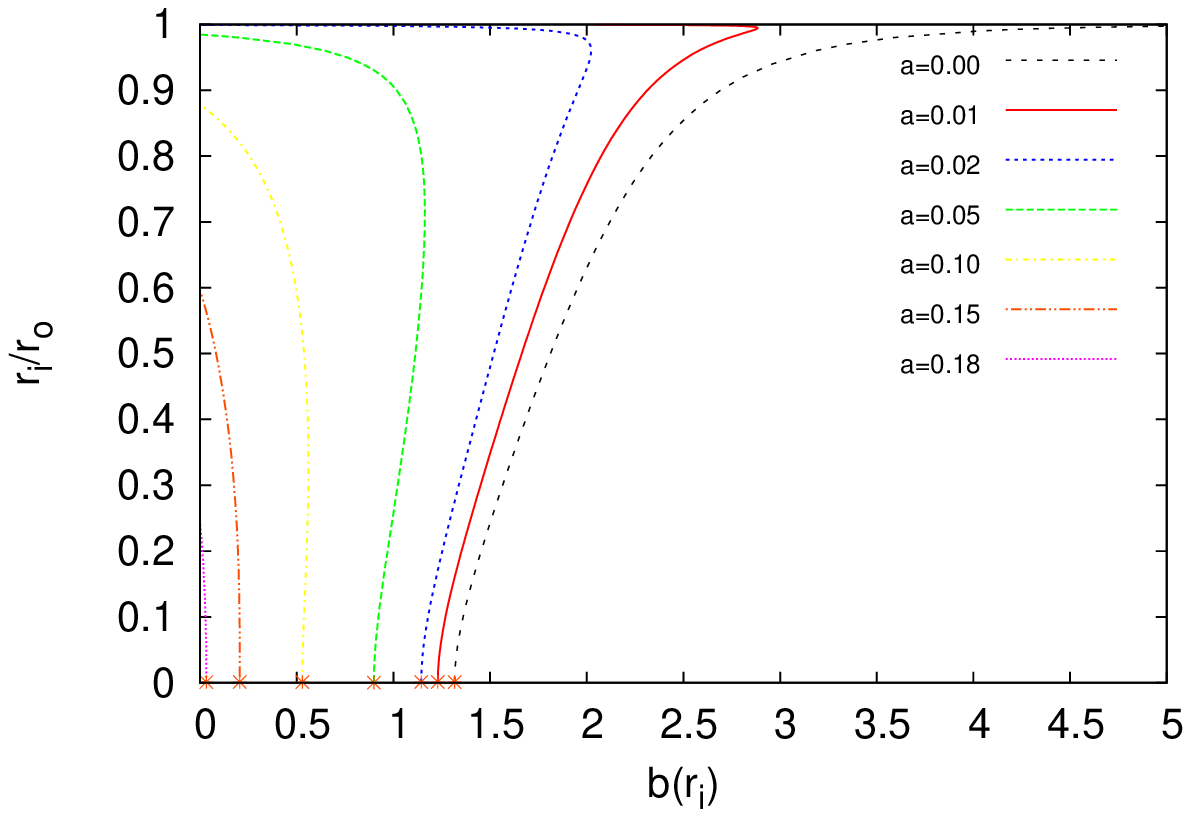}
\label{phaseSdiag2}
}
\subfigure[][]{\hspace{-0.5cm}
\includegraphics[height=.27\textheight, angle =0]{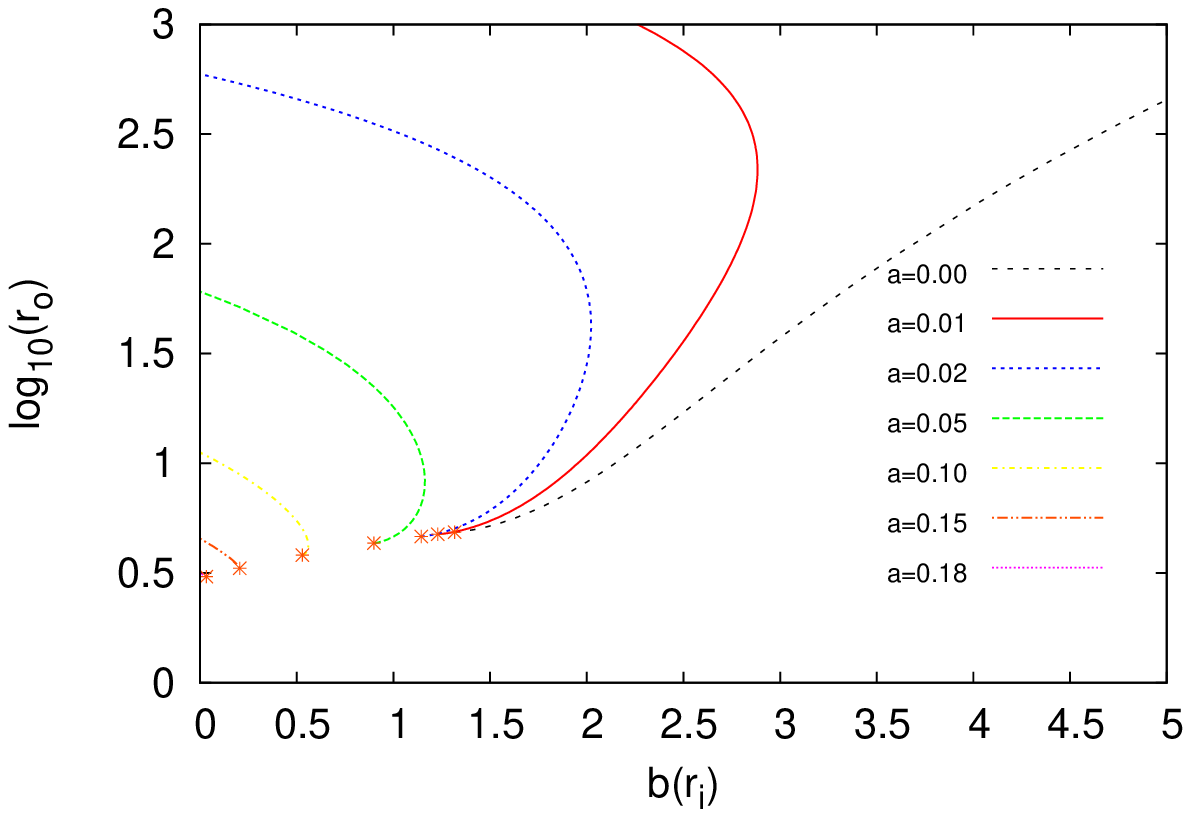}
\label{QS_r0_vs_b1}
}
}
\mbox{\hspace{-0.5cm}
\subfigure[][]{\hspace{-1.0cm}
\includegraphics[height=.27\textheight, angle =0]{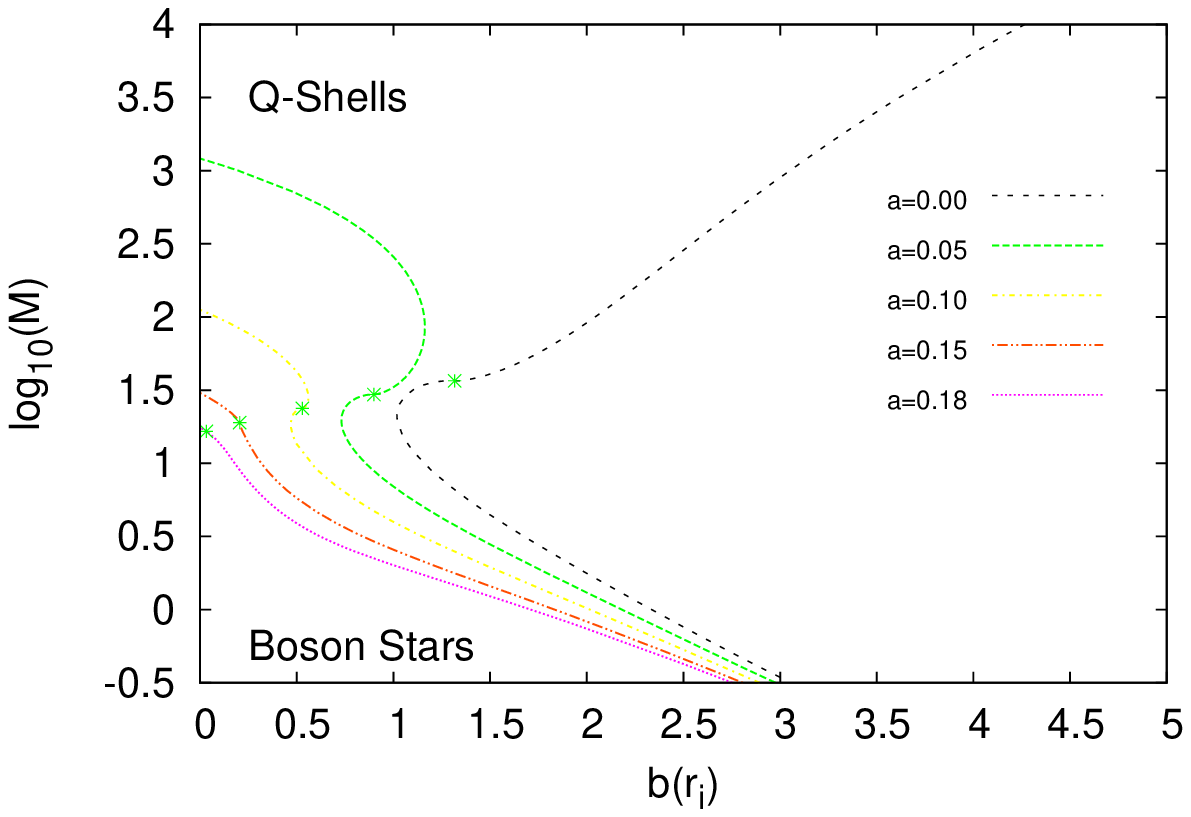}
\label{QBS_M_vs_b1}
}
\subfigure[][]{\hspace{-0.5cm}
\includegraphics[height=.27\textheight, angle =0]{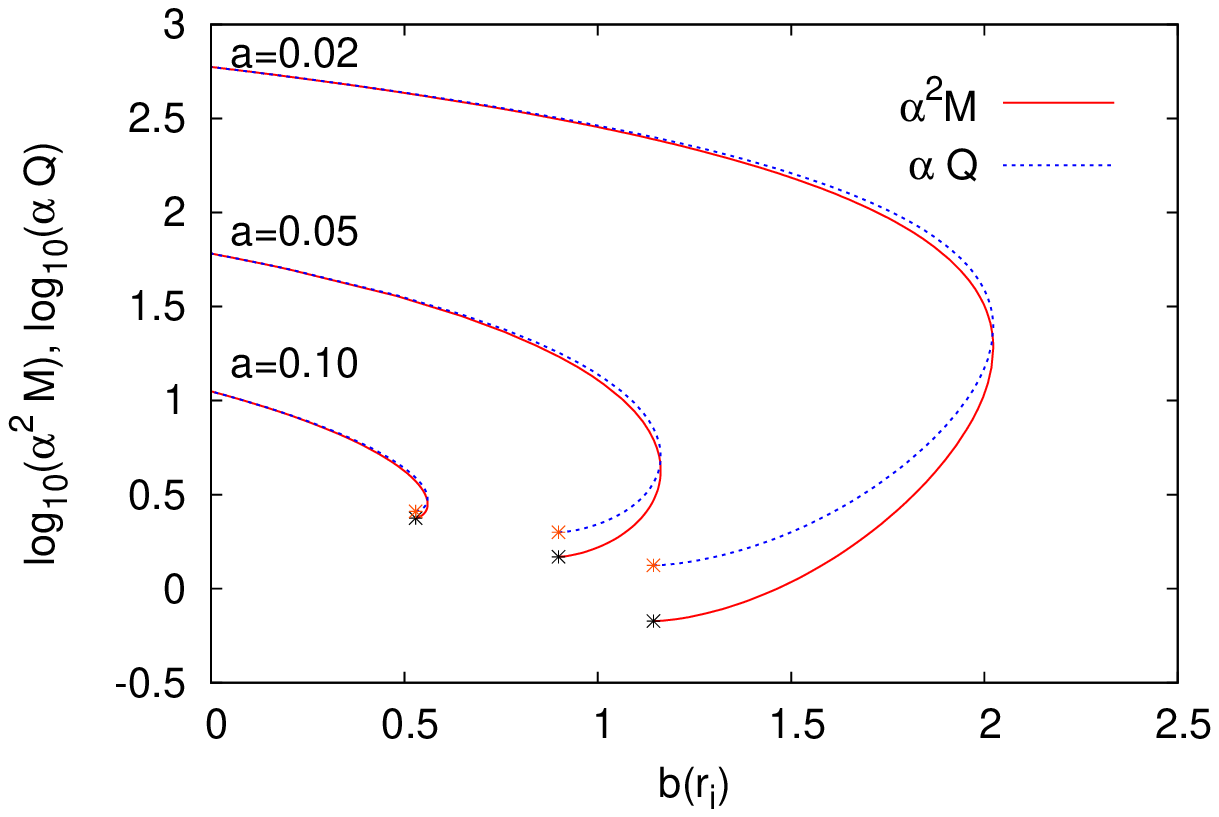}
\label{QS_MQ_vs_b1}
}
}
\end{center}
\caption{Properties of the gravitating $Q$-shell solutions shown versus 
$b(r_i)$,
the value of the gauge field function $b(r)$ at the inner 
shell radius $r_i$:
(a) $r_i/r_o$, the ratio of inner and outer shell radii;
(b) the outer shell radius $r_o$;
(c) the mass $M$ of shell-like solutions and boson stars 
(resp.~$Q$-balls for $\alpha=0$);
(d) the scaled mass $\alpha^2 M$ and the scaled charge $\alpha Q$.
Note that $a=\alpha^2$, and the asterisks mark the transition
points from boson stars ($Q$-balls) to $Q$-shells.
\label{Q_shell}
}
\end{figure}

Let us next consider the gravitating shell-like solutions.
Here the space-time consists of 3 parts.
In the inner part $0  \le r < r_i$ 
the gauge potential is constant and the scalar field vanishes.
Consequently, it is Minkowski-like,
with $N(r)=1$ and $A(r)={\rm const}<1$. 
The middle region $r_i < r < r_o$ then represents the shell of
charged bosonic matter, while the outer region
$r_o < r < \infty$ corresponds to part of a Reissner-Nordstr\"om
space-time, where the gauge field exhibits the standard Coulomb fall-off,
while the scalar field vanishes identically.
This behaviour of the functions is demonstrated in Fig.~\ref{fun} for
the shell-like solution with charge $Q=10$ and 
gravitational coupling constant $\alpha^2=0.1$.

We exhibit in Fig.~\ref{Q_shell} the properties of
shell-like solutions.
Fig.~\ref{phaseSdiag2} shows the ratio of the
inner radius $r_i$ to the outer radius $r_o$ for these solutions.
For a given finite value of the gravitational coupling,
the branch of gravitating shells emerges
at the corresponding boson star solution and ends,
when a throat is formed at the outer radius $r_o$.
As this happens, the value of
$b(r_i)$ reaches zero (or equivalently $b(0)\rightarrow 0$,
since $b(r)$ is constant in the interior, $0 \le r \le r_i$).
The exterior space-time $r > r_o$ then corresponds to the exterior of an
extremal RN space-time.

Thus in contrast to $Q$-shells in flat space, which grow rapidly
in size, mass and charge as the ratio $r_i/r_o \rightarrow 1$,
the growth of gravitating $Q$-shells is limited by gravity,
and the restriction in size, mass and charge is the stronger, 
the greater the value of the gravitational coupling constant $\alpha$.
This is demonstrated in Figs.~\ref{QS_r0_vs_b1}, \ref{QBS_M_vs_b1}
and \ref{QS_MQ_vs_b1}, 
where the outer radius $r_o$, the mass $M$
and the charge $Q$ are exhibited
for a sequence of values of the gravitational coupling constant.

In Fig.~\ref{QBS_M_vs_b1} for comparison also the mass of the
corresponding boson star solutions (resp.~$Q$-ball solutions
for vanishing gravitational coupling constant) are exhibited.
The transitions from the ball-like to the shell-like solutions
are indicated in the figure by the small asterisks.
With increasing $\alpha$ the sets of shell-like solutions
decrease rapidly, until at the critical value
$\alpha_{sh}$ (see Fig.~\ref{phasediag})
they cease to exist.

Fig.~\ref{QS_MQ_vs_b1} exhibits the scaled mass $\alpha^2 M$
and the scaled charge $\alpha Q$ for several sets of
gravitating $Q$-shells. Together with Fig.~\ref{QS_r0_vs_b1}
the figure demonstrates,
that the condition 
for extremal RN solutions,
$r_o= \alpha^2 M = \alpha Q$,
is satisfied
for the shell-like solutions,
as the throat forms at the outer radius $r_o$.

Finally we note, that the shell-like solutions satisfy the mass relation
(\ref{Mreg}) as well.
Consequently the mass relation holds for any two globally regular solutions
of a set with given gravitational coupling constant,
thus relating also ball-like and shell-like solutions.

\section{Black Holes}

\begin{figure}[p!]
\begin{center}
\vspace{-1.0cm}
\mbox{\hspace{-0.5cm}
\subfigure[][]{\hspace{-1.0cm}
\includegraphics[height=.27\textheight, angle =0]{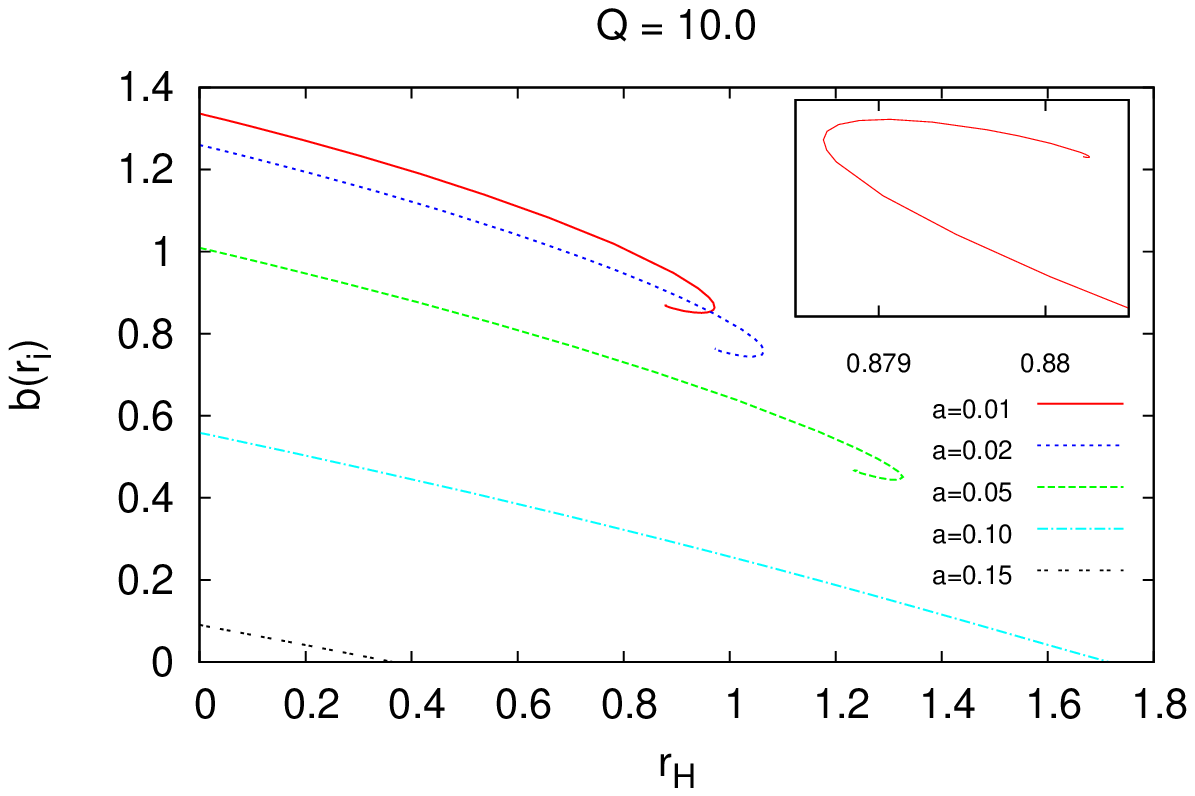}
\label{BHQ10_b1_vs_rh}
}
\subfigure[][]{\hspace{-0.5cm}
\includegraphics[height=.27\textheight, angle =0]{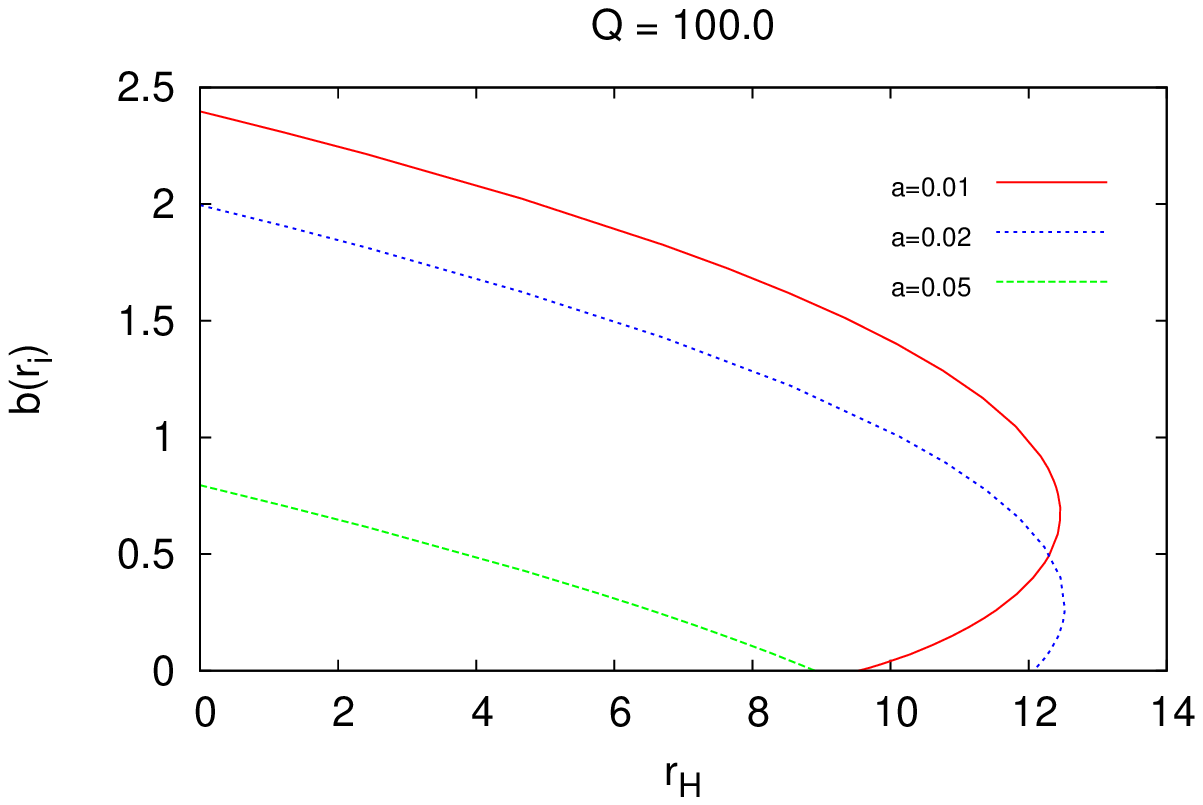}
\label{BHQ100_b1_vs_rh}
}
}
\vspace{-0.5cm}
\mbox{\hspace{-0.5cm}
\subfigure[][]{\hspace{-1.0cm}
\includegraphics[height=.27\textheight, angle =0]{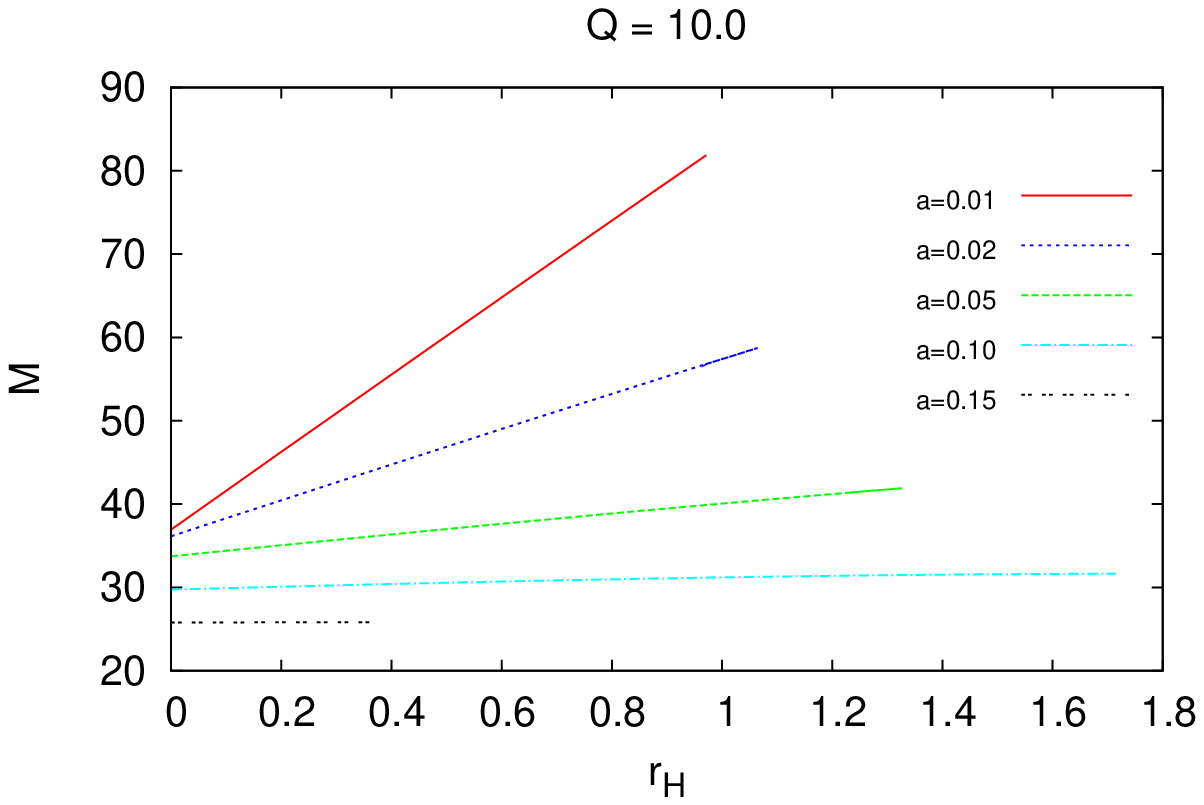}
\label{BHQ10_M_vs_rh}
}
\subfigure[][]{\hspace{-0.5cm}
\includegraphics[height=.27\textheight, angle =0]{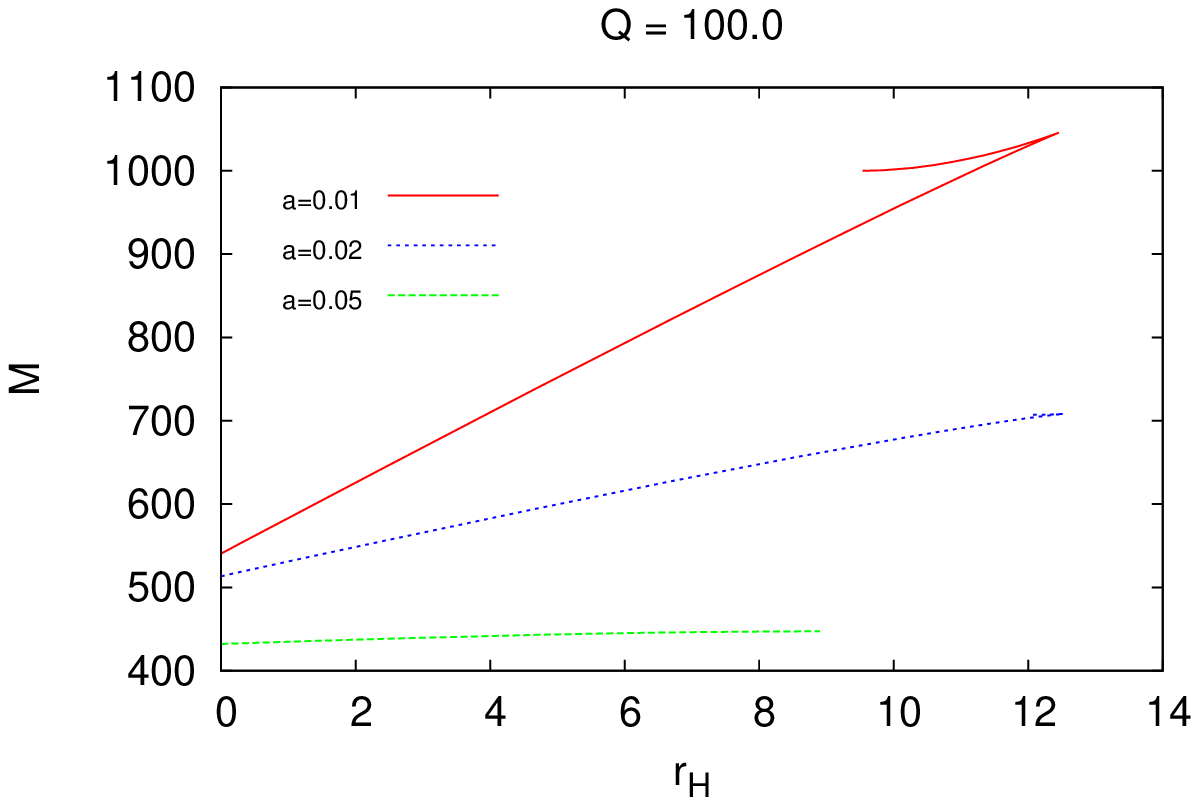}
\label{BHQ100_M_vs_rh}
}
}
\vspace{-0.5cm}
\mbox{\hspace{-0.5cm}
\subfigure[][]{\hspace{-1.0cm}
\includegraphics[height=.27\textheight, angle =0]{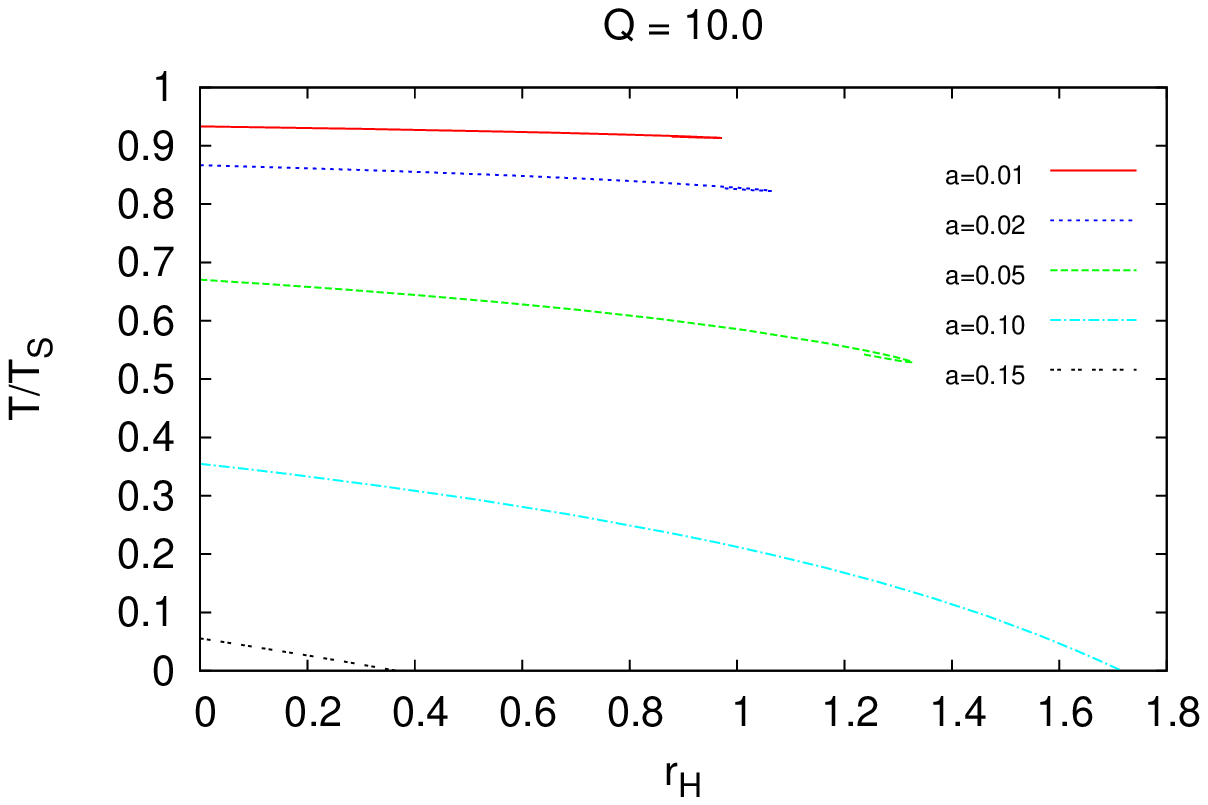}
\label{BHQ10_T_vs_rh}
}
\subfigure[][]{\hspace{-0.5cm}
\includegraphics[height=.27\textheight, angle =0]{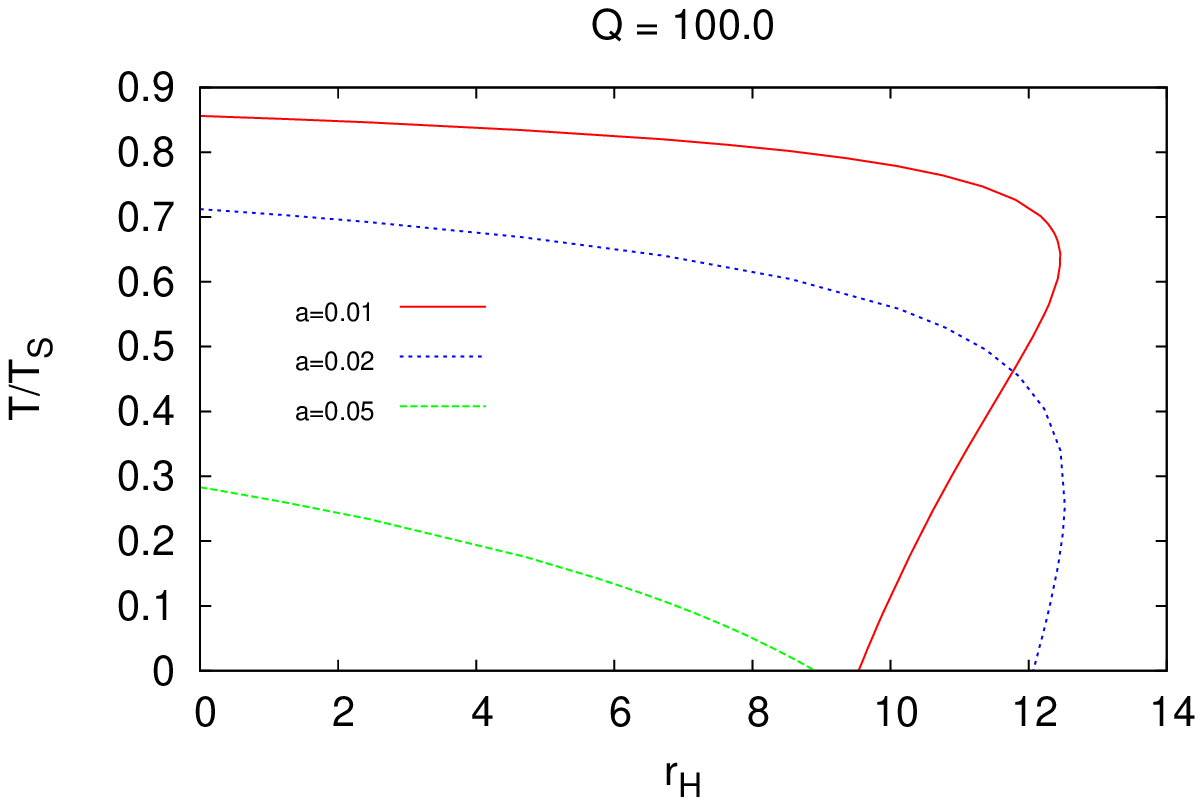}
\label{BHQ100_T_vs_rh}
}
}
\end{center}
\caption{Properties of the black hole solutions 
with Schwarzschild-type interior shown versus 
the horizon radius $r_H$:
The left column exhibits for solutions with fixed charge $Q=10$
(a) 
$b(r_i)$, the value of the gauge
field function $b(r)$ at the inner shell radius $r_i$;
(c) the mass $M$;
(e) the ratio of the temperature $T$ at the black hole horizon $r_H$ to the
corresponding Schwarzschild black hole temperature $T_S$.
The right column ((b), (d), (f)) exhibits the
same for solutions with fixed charge $Q=100$.
Note that $a=\alpha^2$. 
\label{BH1}
}
\end{figure}

Let us finally address black holes in this model.
The simplest type of black holes is obtained,
when the Minkowski-like inner part of the space-time,
$0 \le r \le r_i$, of gravitating $Q$-shell solutions
is replaced by the inner part of a curved Schwarzschild-like space-time.
The metric in the interior region $0 \le r \le r_i$
is then determined by the function $N(r)= 1 - (r_H/r)$ 
and a constant function $A(r)$.
Thus the event horizon resides at $r_H < r_i$.
But the presence of the $Q$-shell outside the event horizon,
makes the properties of the black hole differ from those
of a pure Schwarzschild black hole.

Since with the event horizon size a further variable appears,
which is an important physical quantity,
we discuss the black hole properties with respect to the
horizon radius $r_H$ in the following.
The metric and matter field functions 
for black holes with charge $Q=10$ 
at gravitational coupling $\alpha^2=0.1$
are exhibited in Fig.~\ref{fun} 
for several values of the horizon radius $r_H$.

To illustrate the domain of existence of such black hole solutions,
we again choose a sequence
of values for the gravitational coupling constant,
but we now keep the charge $Q$ fixed, as we vary 
the horizon radius, starting from the corresponding
globally regular $Q$-shell solution.
A respective set of solutions is shown in Fig.~\ref{BH1}
for $Q=$10 and 100.

First of all we note, that the horizon radius is always limited
in size, where the maximal size grows with the charge $Q$.
For small $Q$, e.g.~$Q=$10, we observe two distinct patterns
for the black hole solutions.
The first pattern arises when the
fixed gravitational coupling constant has a value
below a certain critical value.
Here a maximal horizon size is reached,
when the horizon radius $r_H$ gets close to the inner radius 
of the shell $r_i$.
There a bifurcation occurs and a second branch emerges,
which ends at a second bifurcation, where a third branch emerges, etc.
This results in a spiralling pattern,
where the mass $M$ and the temperature $T$ of the solutions tend towards
finite limiting values.
(The first few branches are apparent in Fig.~\ref{BHQ10_b1_vs_rh},
and enlarged in the inlet for a representative value
of the gravitational coupling constant, $\alpha^2=0.01$,
while the higher branches are too small to be resolved there.)

The second pattern is present above that critical value
of the coupling constant. Here the set of black hole solutions
for fixed gravitational coupling ends, when a throat is formed
at the outer shell radius $r_o$.
There the condition for extremal RN solutions,
$r_o= \alpha^2 M = \alpha Q$, is satisfied again,
as seen in Figs.~\ref{BHQ10_M_vs_rh} and \ref{BHQ100_M_vs_rh}.
As the throat forms,
the temperature $T$ at the event horizon of the Schwarzschild-like
black hole $r_H < r_i$ tends to zero,
as seen in Figs.~\ref{BHQ10_T_vs_rh} and \ref{BHQ100_T_vs_rh}.

While appearing at first unexpected,
the reason for the vanishing of the temperature $T$
is the behaviour of the metric function $A(r)$
in $g_{tt}$, since $A(r)$ tends to zero in the interior,
when the throat is formed,
as seen in Fig.~\ref{fun1}.
We recall, that the ratio of the temperature $T$ of
the black hole within the $Q$-shell 
to the temperature 
$T_{\rm S} = (4 \pi r_{\rm H})^{-1}$ 
of the Schwarzschild black hole is given by
$\displaystyle T / T_{\rm S}
  =  \left. r A N' \right|_{r_{\rm H}} = A(r_H)$.

For larger (fixed) values of the charge we always observe
this second pattern,
although the throat may either be reached directly after a monotonic
increase of the horizon radius $r_H$ to its maximum value,
or along a second branch, where the
horizon radius is decreasing again (having passed a bifurcation),
as seen in Fig.~\ref{BHQ100_b1_vs_rh}.

As seen in the figure,
whenever bifurcations occur, there are two (or more) black hole
space-times with the same value of the charge $Q$ and the
same horizon radius $r_H$ (within a certain range of values),
but different values of the total mass $M$
as measured at infinity.
Surprisingly, however, 
there are also two (or more) black hole space-times
with the same value of the charge $Q$ and the same 
value of the total mass $M$ (within a certain range of values).
These black holes thus have the same set of global charges
but are otherwise distinct solutions of the Einstein-matter equations.
Consequently black hole uniqueness does not hold in this model
of scalar electrodynamics.

\begin{figure}[t!]
\begin{center}
\vspace{-0.5cm}
\mbox{\hspace{-0.5cm}
\subfigure[][]{\hspace{-1.0cm}
\includegraphics[height=.27\textheight, angle =0]{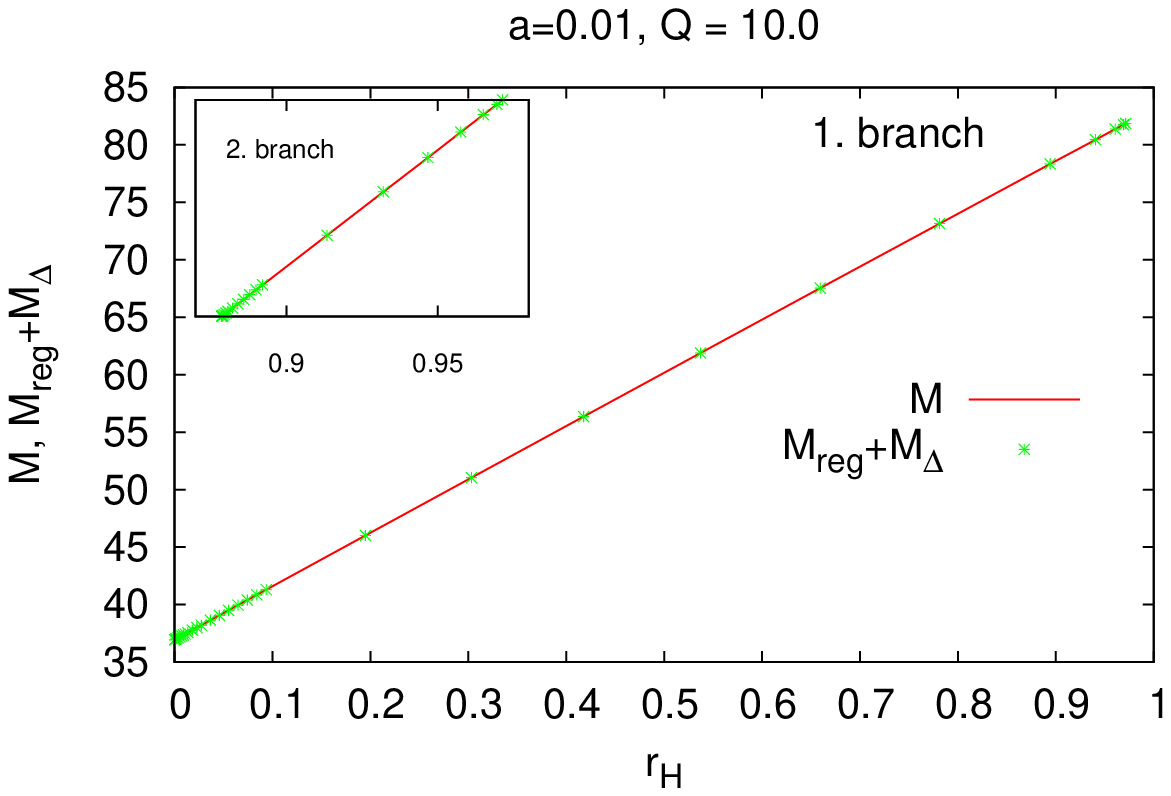}
\label{figiso1}
}
\subfigure[][]{\hspace{-0.5cm}
\includegraphics[height=.27\textheight, angle =0]{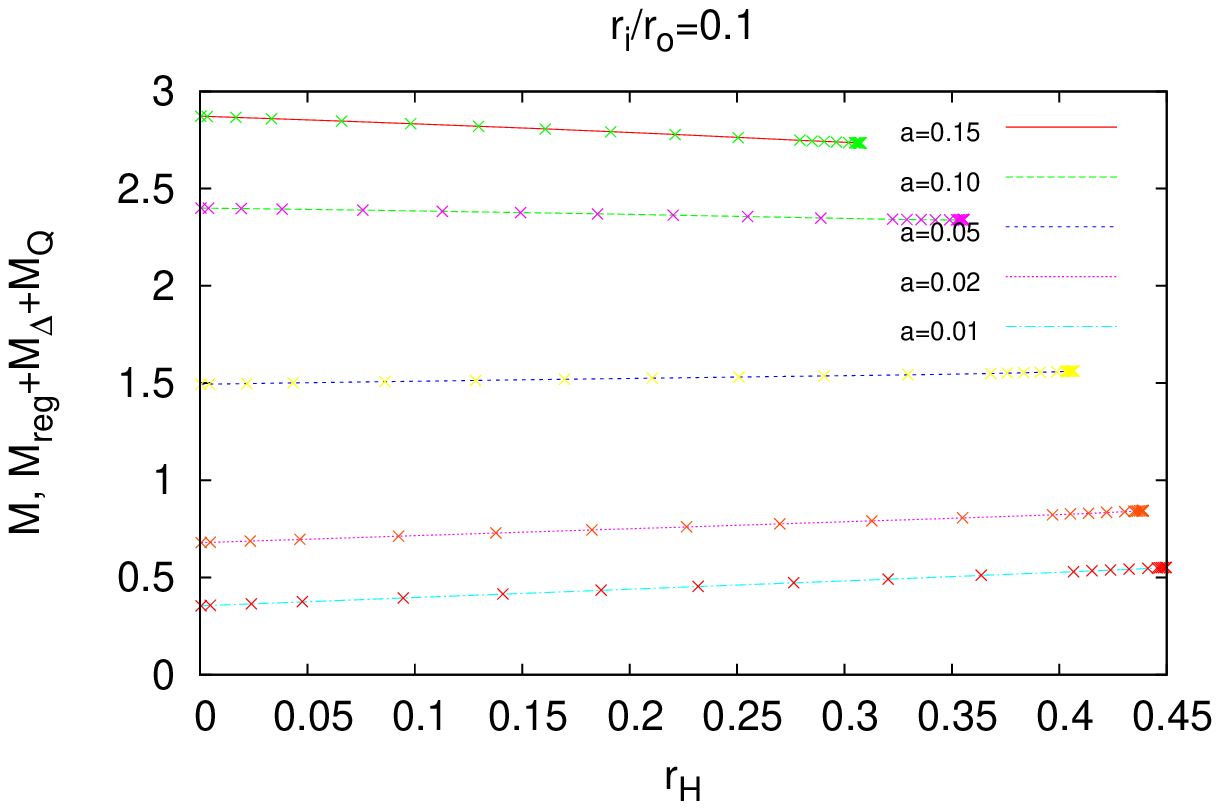}
\label{figiso2}
}
}
\end{center}
\caption{Mass formulae for black hole solutions:
(a) the mass $M$ of solutions with fixed charge $Q=10$ and fixed
gravitational coupling constant $\alpha$,
obtained from the asymptotic metric (\ref{mass}) and the
mass formula (\ref{IHmu});
(b) the mass $M$ of solutions with variable charge $Q$ but
fixed ratio of inner and outer shell radii $r_i/r_o$ for several fixed
values of the gravitational coupling constant $\alpha$,
obtained from the asymptotic metric (\ref{mass}) and the
mass formula (\ref{IHmuDQ2}).
Note that $a=\alpha^2$.
\label{BH2}
}
\end{figure}

Let us finally consider some mass relations
for these black holes space-times possessing $Q$-shells.
We begin by recalling an interesting result
obtained in the isolated horizon framework
\cite{Ashtekar:2004cn}.
It states that the mass $M$ 
of a black hole space-time with horizon radius $r_H$
and the mass $M_{\rm reg}$
of the corresponding globally regular space-time
obtained in the limit $r_H \rightarrow 0$ are related via
\cite{Corichi:1999nw,Ashtekar:2000nx,Ashtekar:2004cn}
\begin{equation}
M = M_{\rm reg} + M_\Delta  ,
\label{IHmu} \end{equation}
where the mass contribution $M_\Delta$ is defined by
\begin{equation}
M_\Delta = \frac{1}{\alpha^2} \, \int_0^{r_H} \kappa(r'_H)r'_H d r'_H  .
\label{IHmuD}
\end{equation}
Here $\kappa(r_H)$ represents the surface gravity 
of the black hole with horizon radius $r_H$,
$\kappa = 2 \pi T$.
Accordingly, the mass $M$ 
of a space-time with a black hole
with horizon radius $r_H$ within a $Q$-shell with total charge $Q$
should be obtained as the sum of the globally regular gravitating $Q$-shell
with charge $Q$ and the integral $M_\Delta$ along the set of black hole
space-times, obtained by increasing the horizon radius for fixed charge
from zero to $r_H$.

This relation is demonstrated in Fig.~\ref{figiso1}
for the set of solutions with charge $Q=10$ and gravitational coupling
constant $\alpha^2=0.01$.
The values for the mass $M$
obtained from the relation (\ref{IHmuD}) are seen to agree with
the values for the black hole mass $M$ obtained from the
asymptotics (\ref{mass}).
The set of solutions exhibited has spiralling character,
i.e., it has besides the main first branch
a second branch, also exhibited, and further branches, 
not resolved in the figure.

When the charge is allowed to vary, too, one expects
a change of the above relation in accordance with (\ref{Mreg2})
and the first law (in the units empoyed), i.e.,
\begin{equation}
dM = \frac{\kappa}{8 \pi \alpha^2} d{\cal A} + b(\infty) dQ
 , \label{firstlaw} \end{equation}
where ${\cal A} = 4 \pi r_H^2$ denotes the area of the horizon
and $b(\infty)$ represents the electrostatic potential at infinity.
Thus we generalize the above relation (\ref{IHmu}) to read
\begin{equation}
M = M_{\rm reg} + M_\Delta + M_Q =
 M_{\rm reg} + M_\Delta 
  +  \int_{Q_{\rm reg}}^{Q} b(\infty) d Q' .
\label{IHmuDQ2}
\end{equation}
This relation is demonstrated in Fig.~\ref{figiso2},
where for several values of the gravitational coupling constant
and for fixed ratio
of inner and outer shell radii $r_i/r_o$,
the values for the mass $M$
obtained from the relation (\ref{IHmuDQ2}) are shown together with
the values for the mass $M$ obtained from the
asymptotics (\ref{mass}).

\section{Conclusion and Outlook}

We have considered boson stars, gravitating $Q$-shells and
black holes within $Q$-shells in scalar electrodynamics
with a $V$-shaped scalar potential,
where the scalar field is finite only in compact ball-like or
shell-like regions.

The gravitating $Q$-shells surround a flat Minkoswki-like interior region,
while their exterior represents part of an exterior
RN space-time.
When the flat interior is replaced by a Schwarzschild-like
interior, black hole space-times result, where
a Schwarzschild-like black hole is surrounded by a compact shell of
charged matter, whose exterior again represents part of an exterior
RN space-time.

These black hole space-times violate black hole uniqueness,
in certain regions of parameter space.
Here for the same values of the mass $M$ and the charge $Q$
two or more distinct solutions of the Einstein-matter equations
exist.

The solutions satisfy certain relations of the type obtained first in the
isolated horizon formalism, which connect the mass $M$ of 
a black hole solution with the mass $M_{\rm reg}$ of the
associated globally regular solution.
The masses of two regular solutions are related in an analogous 
(simpler) manner.
This formalism further suggests to interpret the
black hole space-times as bound states of
Schwarzschild-type black holes and gravitating $Q$-shells
\cite{Ashtekar:2000nx}.

While we have restricted our discussion here to
Schwarzschild-type black holes in the interior,
there are also black hole space-times with charged, i.e.,
Reissner-Nordstr\"om-type interior solutions.
These more general black hole space-times 
will be discussed elsewhere.

The inclusion of rotation presents another interesting generalization
of the solution considered here, 
since rotating boson stars are well-known 
\cite{Mielke:2000mh,Schunck:2003kk,Yoshida:1997qf,Kleihaus:2005me,Kleihaus:2007vk}.
The construction of the corresponding rotating shells
and their black hole generalizations, however, still poses a challenge.

\vspace{0.5cm}
{\bf Acknowledgement}

\noindent
BK gratefully acknowledges support by the DFG,
CL and ML by the DLR.

\end{document}